\documentstyle[psfig,11pt,aaspp4]{article}

\begin{document}
\title{A Photometric Method for Quantifying Asymmetries in Disk Galaxies}
\author{David A. Kornreich\altaffilmark{1}}  
\affil{Center for Radiophysics and Space
Research} \authoraddr{Cornell University Space Sciences Building,
Ithaca NY 14853}
\author{Martha P. Haynes\altaffilmark{1}}  
\affil{Center for Radiophysics and Space
Research and National Astronomy and Ionosphere Center\altaffilmark{2}}
\authoraddr{Cornell University Space Sciences Building,
Ithaca NY 14853}
\and 
\author{R.V.E. Lovelace} \affil{Department of Astronomy,
Cornell University} \authoraddr{Cornell University Space Sciences
Building, Ithaca NY 14853}

\altaffiltext{1}{Visiting Astronomer, Kitt Peak National Observatory
which is part of the National Optical Astronomy Observatories,
operated by the Association of Universities for Research in Astronomy,
Inc. under a cooperative agreement with the National Science
Foundation.}  

\altaffiltext{2}{The National Astronomy and Ionosphere Center is
operated by Cornell University under a cooperative agreement with the
National Science Foundation.}

\begin{abstract}

A photometric method for quantifying deviations from axisymmetry in
optical images of disk galaxies is applied to a sample of 32 face--on
and nearly face--on spirals. The method involves comparing the relative
fluxes contained within trapezoidal sectors arranged symmetrically
about the galaxy center of light, excluding the bulge and/or barred
regions. Such a method has several advantages over others, especially
when quantifying asymmetry in flocculent galaxies.  Specifically, the
averaging of large regions improves the signal--to--noise in the
measurements; the method is not strongly affected by the presence of
spiral arms; and it identifies the kinds of asymmetry that are likely
to be dynamically important. Application of this ``method of sectors''
to {\it R\/}-band images of 32 disk galaxies indicates that about 30\%
of spirals show deviations from axisymmetry at the $5\sigma$ level.

\end{abstract}


\section{Introduction}
Traditionally, studies of spiral galaxies have concentrated on
understanding the dynamics of axisymmetric disks with simple spiral
structure. The high contrast, well--delineated spiral structure in
ideal ``grand design'' spiral galaxies is fairly well understood as
arising from tidal effects.  However, as Baldwin {\it et al.\/}
(1980)\markcite{bls} (hereafter BLS) first pointed out, ideal (that
is, symmetric) galaxies outside of the grand design class are more
uncommon than generally believed. It is now becoming established that
perfect symmetry is not nearly as all--pervasive in either visible
galaxies or in \ion{H}{1} disks as previously assumed.

Asymmetries are generally interpreted (see, for example, Junqueira \&
Combes 1996\markcite{jc}) as the excitation and superposition of $m=1$
spiral modes over the dominant $m=2$ mode, or as local phase--shifting
(with radius) of the $m=2$ mode initiated by gravitational
perturbations due to companion galaxies, even when interacting
galaxies are not in evidence. This idea can be extended to include
those interactions which result in minor mergers, or gas accretion
from the surrounding intergalactic medium, allowing limits to be
placed on the total galaxy accretion rate (Zaritsky \& Rix
1997\markcite{rz97}, hereafter ZR).

Alternatively, asymmetry in the \ion{H}{1} gas velocity field has been
interpreted as an indicator of more subtle dynamics occurring within
the galaxy itself. It is beginning to be understood that stationary
solutions of the hydrodynamics of baryonic and dark matter components
may be non--axisymmetric. Shoenmakers {\it et al.\/}
(1997)\markcite{sfz}, for instance, interpret optical asymmetry as an
indicator of asymmetry in the overall galactic potential, and
therefore an indicator of the (triaxial) spatial distribution of the
dark matter in a galaxy.  Jog (1997)\markcite{jog} has studied the
orbits of stars and gas in a lopsided potential, and reports that
lopsided potentials arising from disks alone are not self--consistent;
rather, a stationary lopsided disk must be responding to asymmetries
in the halo. Finally, some numerical studies such as that done by Syer
\& Tremaine (1996)\markcite{st96} indicate that fluid disks may have
non--axisymmetric equilibrium states, even when embedded in an
axisymmetric halo.

In certain galaxies, NGC~5474 for example (see van der Hulst \&
Huchtmeier 1979\markcite{vdHH} and Rownd {\it et al.\/}
1994\markcite{rownd}), the center of the optical light is separated
from the center of the dynamical \ion{H}{1} distribution, possibly
indicating that galaxies do not rest in stationary states of the
potential at all, but rather that the optical galaxy may be in a
libration about the minimum of the the \ion{H}{1} or dark matter
potential. The preliminary numerical work reported by Levine and
Sparke (1998)\markcite{ls98} is suggestive that optical disks offset
from dynamical centers can result in intriguing forms of
non--axisymmetry.

Several methods have been developed to quantify the frequency of
asymmetry in galaxy images. Rix \& Zaritsky (1995\markcite{rz95},
hereafter RZ) and ZR analyzed the surface brightness distributions in
$K^\prime$-band images of spiral galaxies. Fourier signal strengths
$A_m(r)$ for the $m$ spiral modes were calculated, with large signal
strengths (compared to the average surface brightness $A_0$)
indicating asymmetry. RZ concentrated on the $m=1$ ``lopsided''
Fourier mode as the primary indicator of asymmetry.  Both RZ and ZR
conclude that as many as $30\%$ of galaxies show significant
deviations from axisymmetry by this measure.

Conselice (1997,\markcite{c97} hereafter C97) has proposed that
asymmetries be measured by a method in which galaxy images are rotated
$180\arcdeg$ and subtracted from the original image. Any residuals are
interpreted as asymmetries. C97 does not provide error estimates
of the derived asymmetry parameter, but comments that in the {\it
R\/}-band, at least 54\% of 43 sample galaxies have asymmetry
parameters $A < 0.1$, where $A=0$ and $A=1$ are representative of
maximal symmetry and asymmetry, respectively. The observation is also
made that galaxies tend to be more asymmetric in the {\it J\/}-band
than in the {\it R\/}-band.

Methods for measuring other than global asymmetries have also been
proposed. For example, Elmegreen and Elmegreen (1995)\markcite{ee95}
study the symmetry of the two dominant inner spiral arms in a sample
of 173 bright galaxies. In that study, the relative lengths and
positions of the spiral arms were measured and compared. They find
that in most galaxies, the inner spiral arm lengths are equal to
within 20\%, and are within 20$\arcdeg$ of being 180$\arcdeg$
apart. Such a method could be extended into the outer regions of a
grand design spiral, in lieu of more global methods which can confuse
strong spiral structure with asymmetry.

In their study, BLS compared the \ion{H}{1} surface density profiles
on both sides of the galaxy for a sample of galaxies mapped in
\ion{H}{1} with sufficient resolution to measure the degree of
asymmetry in \ion{H}{1} disks. Using a much larger dataset of global
\ion{H}{1} profiles obtained from the literature, Richter \& Sancisi
(1994\markcite{rs}) classified asymmetry qualitatively. Based
essentially on differences in the fluxes from the two horns, or total
differences in flux between approaching and receding sides, they
assigned galaxies to one of three categories of apparent asymmetry:
``strong,'' ``weak,'' or no asymmetry. Using the qualitative
assignment, Richter \& Sancisi concluded that only 47\% $\pm$ 5\% of
the global profiles show no asymmetry. Furthermore, they argued that
an observed correspondence between obviously asymmetric \ion{H}{1}
maps (such as that for M101) and lopsided line profiles justifies the
proposition that an observed asymmetry in the global \ion{H}{1}
profile can be taken as evidence for a non--circular density
distribution. It should be noted, however, that this is a converse
argument complicated by the fact that \ion{H}{1} profiles contain both
velocity field and \ion{H}{1} distribution information, which are
often difficult to disentangle. Richter \& Sancisi did attempt to
correct for the presence of companions or the presence of unusual
kinematics by eliminating peculiar--looking profiles {\it a priori.\/}
However, because the \ion{H}{1} emission profiles used for this study
were taken from observations made with a smorgasboard of telescopes,
the effects of pointing error (which could manifest itself as an
asymmetry in the profile) could not be determined.

Haynes {\it et al.\/} (1998a)\markcite{h98a} have conducted a similar
study of \ion{H}{1} line profiles in which spectra for a sample of 104
disk galaxies were taken using the 43-meter telescope at Green Bank
and examined for asymmetry in a quantitative way. Using two methods
for the determination of asymmetry, and after determining that
pointing errors produced negligible deviations, the conclusion
supported the previous results that asymmetry is found in $\sim$~50\%
of global \ion{H}{1} profiles. However, those authors caution that
asymmetry in line profiles obtained with single--dish telescopes can
arise for a variety of reasons unrelated to disk asymmetry, such as
contamination from companions within the main beam or sidelobes.

In this paper, we present a new photometric method for quantifying
deviations from axisymmetry in optical images of disk galaxies.
Essentially, we attempt to determine the likelihood that the light 
distribution in a galaxy is inherently symmetric by comparing
the relative fluxes received in different directions from the optical center. 
In \S 2, we present the observational sample of 32 face--on and nearly
face--on galaxies and describe the basic image acquisition and reduction
procedures. In \S 3, the photometric method for the determination
of departures from axisymmetry is introduced and then applied to the
target sample. Finally in \S 4, the method is discussed in comparison
with other quantitative determinations of asymmetry in spiral disks.

\section{Observations}

For the purpose of developing a method of measuring departures from
axisymmetry, deep {\it R\/}-band images were constructed for a set of
32 nearby face--on disk galaxies covering a range of spiral Hubble
classes. In this section, we discuss the observational and image
reduction procedures for the target sample of galaxies.

\subsection{Image Acquisition and Data Reduction\label{reduction}}

The method of sectors described in the following sections was used to
determine differences from symmetry for a sample of 32 face--on disk
galaxies listed in Table \ref{sample} and displayed in Figure
1. Observations were conducted with a Harris {\it R\/}-band filter
under nonphotometric conditions. Each galaxy except NGC~3393,
NGC~3450, and \hbox{MCG-5-34-002} was observed with the KPNO 0.9 meter
telescope configured to $f/7.5$ and equipped with the T2KA direct
imaging camera. The camera setup yielded a gain of 3.6 electrons per
adu and a $0\farcs 68$ per pixel scale. Only the inner $1024\times
1024$ pixels were illuminated, resulting in a field of view of
$11\farcm 6 \times 11\farcm 6$. Read noise in the camera was 4
electrons per pixel. Observations took place during several observing
periods during November and December 1996 and February and March 1997.

The three galaxies NGC~3393, NGC~3450, and \hbox{MCG-5-34-002} were
observed in April 1997 at the CTIO\footnote{The Cerro Tololo
Interamerican Observatory is operated by the Association of
Universities for Research in Astronomy Inc. (AURA), under a
cooperative agreement with the National Science Foundation as part of
the National Optical Astronomy Observatories.} 0.9 meter telescope
with the T1K2 imaging camera. The gain in this camera is 3.35
electrons per adu and the plate scale is $0\farcs 384$ per pixel at
Cassegrain focus. Read noise is 4.6 electrons per pixel.

At both telescopes, individual exposures of 600 seconds each were
taken; the total integration time per object was varied depending on
the atmospheric transparency.

Image reduction was accomplished in the {\it IRAF\footnote{{\it
IRAF\/} (Image Reduction and Analysis Facility) is distributed by the
National Optical Astronomy Observatories.}\/} image processing
environment. These exposures were then debiased, sky--subtracted, and
flatfielded using twilight flats. The exposures were scaled by
measuring the integrated fluxes from 5--10 stars in the field. Because
the transparency was variable, the frames were scaled to the single
frame showing the greatest number of counts regardless of airmass. The
frames were then combined using weights inversely proportional to the
scale factor used, to produce images with total exposure times of
between 3000 and 4800 seconds. Because the images are not photometric
(and in some cases, far from photometric), the relation between
exposure times and total integration time is not clear. This
variability does not affect the measurements conducted here, however,
because the interesting quantities involve only magnitude differences
of regions within individual images. Estimated errors which involve
the integration time, however, may be affected.

After combination, foreground stars were removed from the frame and
replaced with a second--order polynomial interpolation of the
surrounding background. A record of the number of pixels so replaced
was kept, to be used as part of the final error estimate. Elliptical
isophotes were then fit to the galaxy using a slightly modified
version of the STSDAS\footnote{STSDAS (Space Telescope Science Data
Analysis System) is distributed by the Space Telescope Science
Institute which is operated by AURA under contract to the National
Aeronautics and Space Administration.} ISOPHOTE package (Haynes {\it
et al.\/} 1998b\markcite{h98b}). The model galaxy thus obtained was
used to determine each galaxy's center of light, inclination, and disk
scale length $R_d$. Scale length was determined by finding the best
fit line in a surface brightness profile for the outer regions of a
galaxy. Inclination was estimated under the assumption that galaxies
seen face--on should appear circular.  This assumption, although widely
used to determine inclinations, is suspect in a project wherein
inherent circular symmetry is not assumed. It is, however, justified
statistically by studies such as that by Binney and de Vaucouleurs
(1981)\markcite{bdv} and in RZ, in which the observed distribution of
galaxy ellipticities on the sky was used to estimate the distribution
of the intrinsic axial ratios of disk and galaxies. We thus take the
intrinsic ellipticities of disk galaxies as finite, but well within
the errors of our ellipticity measurements. These values and the
formal numerical errors are included in Table \ref{sample}. Following
this basic data reduction, images were analyzed for asymmetry using
the method of sectors described in the next section.

The images shown in Figure \ref{allgals} are individually scaled and
trimmed for illustrative purposes, and most stars therein are masked
to convey the visual appearance of each object.

\subsection{The Sample}

The observational sample selected for this study was chosen based on
applicability for imaging for the purpose of employing the method of
sectors discussed in \S \ref{pizzasection}. The primary criterion was
face--on aspect, to eliminate confusion due to, for instance, the
presence of dust lanes and extinction in the disk; the main source was
the {\it Third Reference Catalogue of Bright Galaxies\/} (de
Vaucouleurs {\it et al.\/} 1976\markcite{rc3}; RC3).  The galaxies in
the observed sample were selected based on apparent disk size $1\farcm
5 < D_{25} < 5\arcmin$, apparent ellipticity $e^* = R_{25}-1 < 0.15$,
and narrow ($W_{HI} \lesssim 100$~km~s$^{-1}$) \ion{H}{1} line
width. The latter restriction is a further diagnostic for low
inclination in rotation--dominated disks. Since turbulence contributes
to broadening at a level of $\sigma_V$ (Broeils 1992\markcite{b92};
Giovanelli {\it et al.\/} 1998\markcite{rg98}) $\sim 8-15$~km~s$^{-1}$
and deviations from coplanar rotation may be quite common (Lewis
1987\markcite{l97}), \ion{H}{1} line widths arising from spirals that
are observed to be as narrow as 100~km~s$^{-1}$ clearly suggest that
the disk is seen face--on.  Figure \ref{HIwidths} demonstrates how
rarely \ion{H}{1} line widths are observed to be so narrow. The top
panels shows the distribution of observed \ion{H}{1} line widths
W$_{HI}$ for a sample of 3123 galaxies in the Local Supercluster for
which \ion{H}{1} line widths are available either in the RC3 or,
preferentially, in our own private database.  Reported velocity widths
have been converted to a systematic definition and corrected for
instrumental broadening, signal--to--noise and redshift stretch,
following Giovanelli {\it et al.\/} (1998) and Haynes {\it et al.\/}
(1998c)\markcite{h98c}.  The lower three panels show the distributions
for spiral subtypes as indicated. Most of the narrow W$_{HI}$ profiles
seen in the total sample arise from the low mass dwarf irregular
galaxies, whereas only a small percent of spirals, those with
inclinations $\le 15^{\circ}$, have narrow widths.  We have thus used
such widths as an important indicator of low inclination.

Because of the desire to understand the nature of departures from
symmetry in isolated disks, galaxies with no known nearby companions
were preferentially included in the sample. In contrast, NGC~5474,
which lies within the gravitational potential of M101 and is
well--known for its high degree of asymmetry was included for
comparative purposes. Likewise, NGC~1637 was added to the sample
because of its well--known optical and infrared asymmetry despite its
symmetric \ion{H}{1} line profile (see Haynes {\it et al.\/} 1998a and
Block {\it et al.\/} 1994\markcite{block}).

Because the observations were conducted during non--photometric
conditions when the main program could not be carried out, final
target selection was not controlled but rather dependent on the
ill--fortune of the primary program. Table \ref{sample} lists relevant
information about the program objects.  Entries in the table are as
follows:

\noindent
Col. 1: Entry number in the Uppsala General Catalog (Nilson
1973)\markcite{ugc}, where applicable, or else in our private
database, referred to as the Arecibo General Catalog (AGC).

\noindent 
Col 2: NGC or IC designation, or other name, typically from the
Catalog of Galaxies and Clusters of Galaxies (Zwicky {\it et al.\/}
1960)\markcite{cgcg}, the ESO--Uppsala Catalog (Lauberts
1982)\markcite{EU} or the Morphological Catalog of Galaxies
(Vorontsov--Velyaminov \& Arhipova 1968)\markcite{mcg}.

\noindent
Cols. 3 and 4: Right Ascension and Declination in the 1950.0 epoch,
typically from the AGC. In general, the listed positions have 15" accuracy.

\noindent
Col. 5: The morphological type code from the RC3.

\noindent
Col. 6: The major and minor diameters, $D_{25}$ and $d_{25}$, in arcmin, 
from the RC3.

\noindent
Col. 7: The \ion{H}{1} line width, W$_{HI}$, corrected for instrumental broadening,
smoothing, and signal--to--noise, in km~s$^{-1}$. Values are taken from the 
literature as recorded in the AGC
and have been converted with appropriate corrections
to the system adopted by Haynes {\it et al.\/} (1998c) and
Giovanelli {\it et al.\/} (1998\markcite{rg98}).

\noindent
Col. 8: The inclination and its associated error, in degrees, derived
from the ellipse fitting procedure described in \S \ref{reduction},
under the assumption of an infinitely thin disk. The errors presented
are formal numerical errors derived from constraints on geometrical
quantities associated with the isophotes, including position angle and
ellipticity. Because the galaxies are so nearly face--on,
uncertainties in position angle are typically large. We therefore
believe that the presented errors are overestimates.

\noindent
Col. 9: The heliocentric systemic velocity, $V_{\odot}$, in km~s$^{-1}$,
from the AGC.

\noindent
Col. 10: The total exposure time of the combined {\it R\/}-band image.
Note that the images were all obtained under non--photometric
conditions of differing transparency. As discussed in \S
\ref{reduction}, some compensation for the variation in conditions was
made by varying the total exposure times and by weighting each
exposure in forming the combination by the inverse of its scale
factor.

While the sample size is too small to treat its completeness, it was
compared to the larger sample of all spiral galaxies of distance $cz <
3000$~km~s~$^{-1}$ in the RC3 with observed values of $D_{25}$ and
$R_{25}$, with the subsample thereof including only those ``face--on''
galaxies with $R_{25} < 1.3$, and with the subsample comprised of
those face--on galaxies within 2000~km~s$^{-1}$.  The distributions of
several parameters were examined for each sample and mean values
obtained. For a thorough investigation into the completeness of the
catalogue, and distributions of parameters in the Local Supercluster,
see Roberts \& Haynes (1994)\markcite{rh94}. The parameters examined
were: the apparent optical diameter $D_{25}$, the apparent axial ratio
$R_{25}$, corrected bolometric magnitude $B_T^\circ$, \bv and \ub
~colors, mean surface brightness $m_e^\prime$, heliocentric velocity,
and type index. The analysis indicates that the sample galaxies, with
$\left< D_{25} \right> = 3\farcm 66 \pm 0\farcm 9$, are larger on the
sky than on average for the supercluster $\left( \left<D_{25}\right> =
2\farcm 12 \pm 1\farcm 0 \right)$. The mean \bv and \ub colors and
surface brightness of the sample are those of the local supercluster
to within errors. The sample spans the spiral sequence with a mean
type of Sc. While the sample is in no way complete, it is nonetheless
representative of face--on galaxies within the Local Supercluster.

\section{The Method of Sectors\label{pizzasection}}
We wish to construct a quantitative method for determining when a
galaxy disk differs significantly from axisymmetry. We have decided on
a method which is geometrically based, provides a global measure of
asymmetry, and is applicable to a wide variety of data sets.

The chosen method quantifies departures from disk axisymmetry by
dividing the galaxy into a number of equal--area trapezoidal ``wedges''
in which photometry is to be performed (see Figure
\ref{pizzapie}). The wedges are derived from a series of triangles,
with the apex of each located at the center of light of the
galaxy. The lateral boundaries of each wedge are radials emanating
from this center. Each triangle is then truncated at a predetermined
radius near the bulge, and extended outwards to the edge of the
visible galaxy, thus creating a trapezoidal section. For most galaxies
in the sample, inner and outer boundaries of $R_d$ and $5R_d$ were
chosen to ensure that the analysis would not be affected by the bulge
component, especially including asymmetries introduced by bars, while
including as much of the disk luminosity as possible. When the galaxy
under consideration is slightly inclined, the sector pattern can be
projected to the appropriate angle of inclination, generally
$15^\circ$ or $30^\circ$. Photometry is conducted in each of the
sectors, and the magnitude of each sector is compared to that of the
others. The magnitude difference between sectors $i$ and $j$ is
$\Delta M_n^{ij} \equiv \left| M_i - M_j\right|$, and the largest
magnitude difference between sectors is $\Delta M_n^{max}$, where $n$
represents the number of sectors used. $\Delta M_n^{max}$ is the
quantitative measure of asymmetry, and where this quantity is
significantly different from zero, this is taken as a detection of
such.

To ensure that the method was sensitive to asymmetries in the visible
flux, the method was tested by considering the very asymmetric galaxy
NGC~5474 (Figure \ref{n5474}). This galaxy contains a nucleus which is
so offset from the disk structure as to be disconnected from it. It
does, however, have a roughly north--south line of approximate
reflection symmetry. Using 6 sectors in a configuration aligned with
the line of symmetry, we performed photometry in the sectors and
determined the greatest difference between segments was $\Delta
M_6^{max}=2\fm 14$ magnitudes. Because the sectors were aligned with
the line of symmetry, we expected complimentary sectors to have
similar total magnitudes. In this test, complimentary sectors across
the line of symmetry differed only by an average of $\Delta M^c_6=0\fm
15$ magnitudes. Similar results were obtained when the sector
boundaries were allowed to vary away from perfect alignment with the
symmetry axis by $5\arcdeg$, and for sets of 8 and 10 sectors, with
average complimentary sector differences being $\Delta M^c_8 = 0\fm
20$ and $\Delta M^c_{10}=0\fm 23$, respectively. The maximum magnitude
differences in these cases were $M_8^{max} = 2\fm 34$ and
$M_{10}^{max} = 2\fm 36$, consistent with $\Delta M_6^{max}$ to within
the values of $\Delta M^c$.

Errors in the measured flux of a region were estimated as the sum of
two distinct factors. The first of these is the formal measurement
error in the flux $\delta\sigma$ including both the expected read
noise and Gaussian sky noise. The second factor represents
uncertainties in the masking of foreground stars. These two errors
were added in quadrature for each galaxy to obtain an error estimate
in the flux, and then transformed via the usual relations to a final
error estimate in the magnitudes.

The most ideal measure of the errors associated with this method would
be to perform photometry on the sky in each frame, using the same
sector pattern as used on the galaxy. The image frames, however, were
generally not large enough to permit this.  In order to understand the
magnitude of the errors involved, therefore, the method of sectors was
applied to a symmetric calibrator galaxy.  For this calibration, it
was important to choose a symmetric, flocculent disk galaxy so that
neither spiral structure nor the presence of strong \ion{H}{2} regions
would contribute disproportionate amounts of luminosity to any one
wedge. For these reasons, the galaxy \hbox{MCG-1-60-011} was
selected. Although classified as Sd in the RC3, this galaxy exhibits
rather tightly wound, very symmetric flocculent spiral structure and
exhibits few \ion{H}{2} regions.

Observed values of the maximum flux difference $\Delta f^{max}$ for
\hbox{MCG-1-60-011} were interpreted as $1\sigma$ errors away from perfect
symmetry in the fluxes. Furthermore, formal errors $\delta\sigma$ were
calculated for each flux determination. If $\rho_{\rm
mask}$ is the density of masked pixels in the image, and $A$ is the
area of a sector, then the errors associated with masking should be
proportional to $A\rho_{\rm mask}$. We therefore introduce the
proportionality constant $\beta$, such that when added in quadrature,
the total error estimate becomes 

\begin{equation}
\sigma = \sqrt{\delta\sigma^2 + \beta^2A^2\rho_{\rm mask}^2}.
\end{equation}

If the $\Delta f^{max}$ values obtained for this galaxy are truly
representative of symmetry, then we expect them to be of order
$\sigma$.  Setting $\sigma = \Delta f^{max}$ {\it a priori,\/} it is
possible to obtain a value of $\beta$ for each segment size used. We
find $\beta$ to be roughly constant over segment size, thus the
proposition of using $\Delta f^{max}$ for the estimates of the errors
is justified. A value of $\beta = 92.3$~ADU-pixel was taken for use
with the remainder of the sample, as this is the average of the three
values, none of which vary from the mean value by more than 5\%.

Table \ref{results} contains the measured values of $\Delta M^{max}$
for the 32 galaxies in the sample with $1\sigma$ errors. Each galaxy
was studied using three segment patterns consisting of 6, 8, and 10
sectors. Using a larger number of segments effectively compares
smaller portions of the galaxy than when fewer segments are used, but
in this regime, the estimated errors tend to increase. For the
galaxies more than $5\sigma$ away from symmetric, we have included the
type of asymmetry seen based on the location of the minimum and
maximum flux segments in the 8-segment runs. When the extreme segments
are adjacent to one another, we desginate the galaxy ``bisymmetric.''
When the extreme segments form a 135\arcdeg or 180\arcdeg angle, we
call the galaxy ``lopsided.'' Otherwise, the extreme segments form a
90\arcdeg angle and the galaxy is ``boxy.''

Of a total of 32 galaxies, 10 (31\%) are asymmetric at the $5\sigma$
level, while 19 (59\%) are asymmetric at the $3\sigma$ level, where
``asymmetric'' is taken to mean $\Delta M^{max} > 0$ at the stated
confidence level in at least one trial. Because the values of $\Delta
M^{\rm max}$ are positive definite, however, it is important to note
that any given measurement is more likely to be an overestimate than
an underestimate of the ``true'' asymmetry. We therefore prefer to use
the $5\sigma$ values to report that approximately 30\% of nearby field
galaxies are optically asymmetric. This result agrees with the
estimate found by RZ and ZR.

Standard error estimates as described above could not be applied to
NGC~5474, because the galaxy is so asymmetric that all attempts at
isophotal ellipse fitting failed. Due to the similarities between the
error estimates obtained with \hbox{MCG-1-60-011} and the $\Delta M^c$
values obtained for NGC~5474, we adopted $\Delta M^c_n$ as the error
estimates for that galaxy. If anything, these are probably
overestimates.

\section{Discussion}

As illustrated in Table \ref{results}, the degree of asymmetry
returned by the method of sectors agrees roughly with a qualitative
eyeball judgment of asymmetry for each galaxy. The only truly deviant
result was obtained from NGC~1073, which while qualitatively only
slightly asymmetric, returns a very significant ($\simeq 28\sigma$)
deviation from symmetry. We believe that this result is due the fact
that although the galaxy appears relatively symmetric in the disk, the
nucleus is offset from the center of the bar by about 10$\arcsec$ to
the northeast. It should be noted also that this galaxy exhibits a
separate region of greatly enhanced surface brightness along the
bar. Although this region did not formally affect the measurements, it
emphasizes the fact that the bar region of NGC~1073 is peculiar.  Such
an offset of the nuclear region manifests itself as a strong lopsided
asymmetry.

In order to determine whether asymmetric galaxies share any other
common characteristics, the subsample of the nine galaxies found to
differ from symmetry by at least $5\sigma$ (except for NGC~1073 and
the peculiar galaxy NGC~5474) was compared to the overall sample of 32
galaxies.  If we consider SAB galaxies to have .5 of a bar, then the
average ``barredness'' of the asymmetric galaxies is about 0.44,
compared to 0.53, the average ``barredness'' of the overall sample. We
therefore find that lack of symmetry is about evenly distributed
between barred and unbarred galaxies. The average morphological type
of the asymmetric sample is Sbc.

We also report on the dispositions of the five galaxies which were
studied both by ZR and the current work. These galaxies are: NGC~600,
NGC~991, NGC~1302, NGC~7742, and \hbox{MCG-5-34-002} (also known as
ESO~446~G~031). Of these five galaxies, RZ and ZR find that all but
NGC~991 and \hbox{MCG-5-34-002} have values of $\left<A_1\right> <
0.2$, which they interpret as indicative of overall
symmetry. \hbox{MCG-5-34-002} is reported to have a value
$\left<A_1\right> = 0.210 \pm 0.021$. As stated in ZR,
$\left<A_1\right>$ measurements are biased upwards, especially for
small $\left<A_1\right>$, so this measurement is not inconsistent with
symmetry. These findings agree with the current work, in which only
NGC~991 is found to be different from symmetric by more than
$5\sigma$.

We also examine the galaxies in both our sample and the sample of
C97. These galaxies are: NGC~3596, NGC~3631, NGC~4136,
and NGC~5701, which in the latter study are given {\it R\/}-band
asymmetry parameters of 0.10, 0.12, 0.14, and 0.054
respectively. Although C97 does not provide a ``cut--off'' value
for the asymmetry parameter $A(R)$, we can see in that study that
values range from $A(R)=0$ for the most symmetric galaxies to about
$A(R)=0.16$ for the most asymmetric. Our data are roughly consistent
with that of C97, in that we find that the former three galaxies
are asymmetric and the latter is symmetric.

It is interesting to note also that global asymmetry may not
necessarily be correlated with asymmetry in the inner regions, as
measured by Elmegreen and Elmegreen (1995). Even the strongly
asymmetric galaxy NGC~1637 is reported to have an inner arm length
difference of only 23\%, and an arm position difference of only
20$\arcdeg$, a result only slightly above the median in both cases.

The method of sectors has advantages over other methods in that it is
neither dependent on assumed dominant even spiral modes, nor is it
particularly sensitive to small deviations in inclination, which may
manifest themselves in the magnitude of the $m=2$ mode. This is not
only because the sectors can be modified to account for slight
inclinations, but also because even if the sectors are not modified, a
slight deviation $\delta i$ in inclination will be detected as a
slight decrease in the magnitude of sectors along the minor axis
only of order $2.5\log\left(\cos\left(\delta i \right)\right)$. This
error is negligible if $2.5\log\left(\sin\left(\delta i \right)\right)
< 0\fm 1$ for most galaxies, requiring $\delta i \lesssim
25\arcdeg$. This method also, because of its purely geometric nature,
is well suited to find asymmetries other than the $m=1$ ``lopsided''
type. Galaxies shaped like dumbbells, or with ``boxy'' or
``triangular'' shapes, which although asymmetric may be cut into
equally luminous halves, would also be detected with this method.

The method of sectors is suited both to the study of flocculent
galaxies, and of galaxies with strong spiral arms, so long as those
arms extend around the entire galaxy. This is because the integration
over large sections tends to ``wash out'' effects of the spiral arms
which might be seen as asymmetries in the Fourier method of RZ, for
instance. We believe that this is a desirable effect, since we would
not wish to detect symmetric spiral structure as an asymmetry. 

The method of sectors is also applicable to cases where the
radially-dependant Fourier method is inapplicable or fails. The
philosophy of the methods of RZ and C97 seems to be that the only
truly symmetric galaxy is one where the surface brightness is a
function of $r$ alone. We believe that this definition ignores the
central characteristic of disk galaxies, i.e. that they possess spiral
arms, and even $m=1$ spirals are not in general considered by
observers to be qualitatively asymmetric.  Thus these two methods
would misclassify obviously symmetric, but predominantly one--armed
spiral galaxies such as NGC~2326 (Figure \ref{n2326}) as lopsided. We
therefore prefer to define symmetry based on equality of total
integrated flux in various directions outward from the bulge. The
method of sectors is designed to distinguish exactly such symmetry.

Moreover, while the Fourier method is relatively easily applied to
optical data from large nearby galaxies, it becomes progressively
buried in noise as the data become more scarce.  Such a method is not
likely to be applicable to, for example: distant galaxies in the
optical, HI synthesis data, or output of numerical modeling.  The
sector method does not suffer nearly as much from undersampling and
allows direct comparison of optical photometry to other less prolific
data sets. Often, numerical work is presented ``as is,'' with few
quantitative measures available to compare numerical results to actual
data. In this regard, because of its use of flux integration over
large regions, the method of sectors is easily adaptable to
measurements of computer--generated data, allowing direct,
quantitative comparison to natural phenomena. In the future, this
method will be applied to aperture synthesis \ion{H}{1} line mapping
data and the results of numerical simulations of a subsample of the
galaxies observed here.

\acknowledgments 

This research has been partially supported by NSF grant
\hbox{AST95--28860} to MPH, and \hbox{AST93--20068} to RVEL, and has
made use of the NASA/IPAC Extragalactic Database (NED) which is
operated by the Jet Propulsion Laboratory, California Institute of
Technology, under contract with the National Aeronautics and Space
Administration. The authors would also like to thank Daniel Dale for
acquiring the images obtained at CTIO.

\newpage

\begin{table}
\dummytable\label{sample}
\end{table}
\psfig{figure=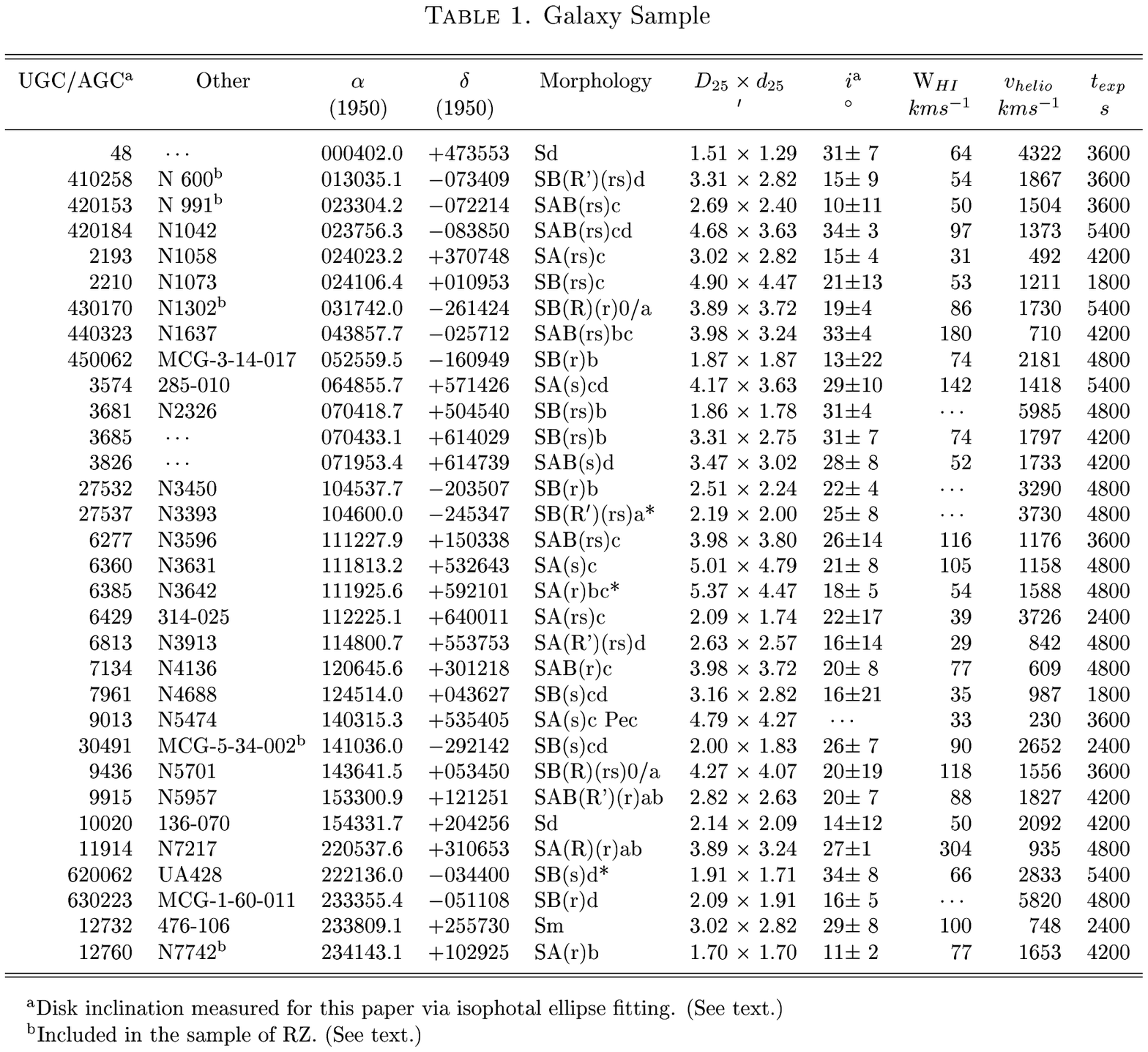,width=7in}

\begin{table}
\dummytable\label{results}
\end{table}
\psfig{figure=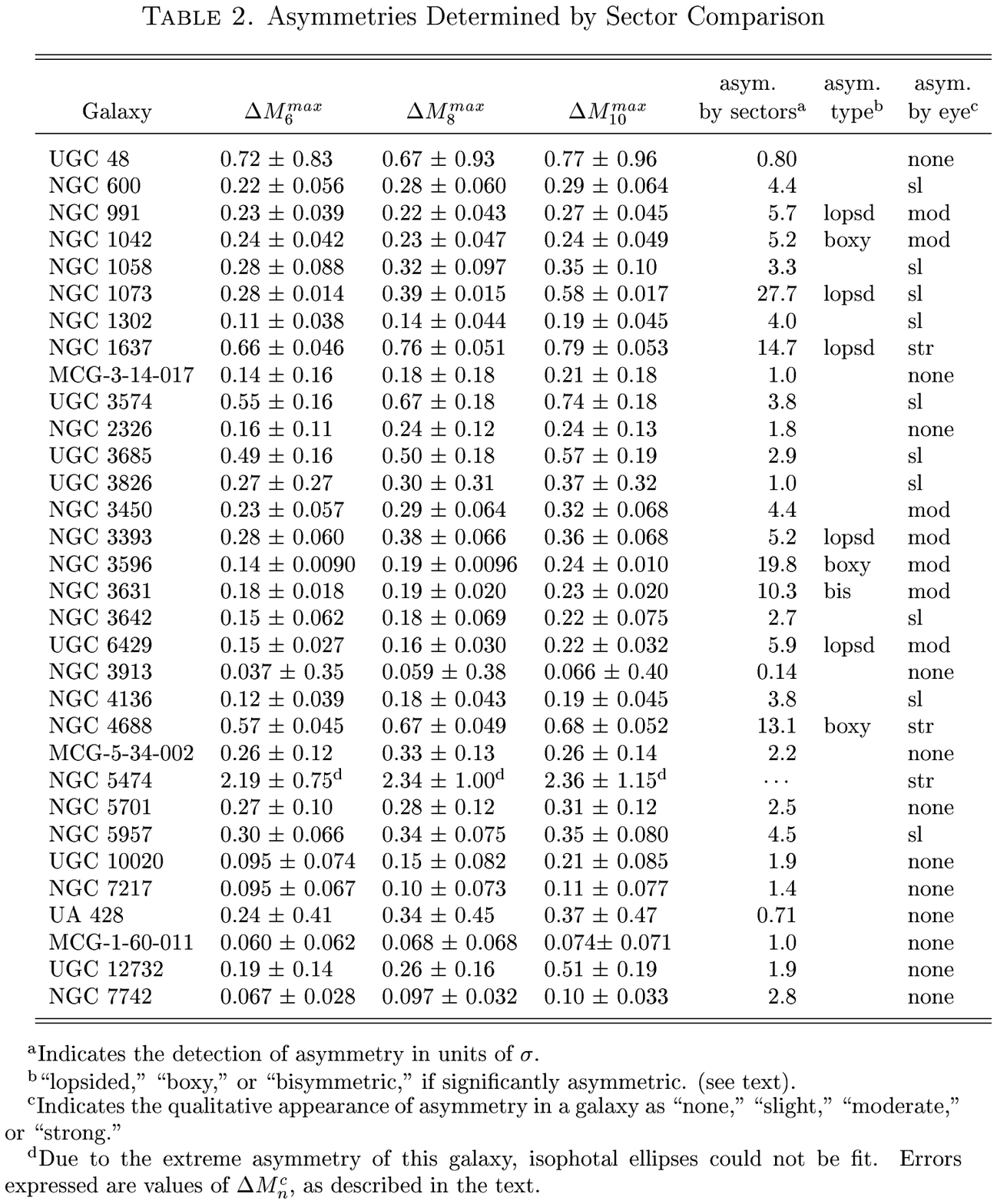,width=7in}

\newpage

\begin{figure}
\figurenum{0}
\end{figure}

\begin{figure}
\figurenum{1}
\caption{Illustration of the 32 galaxies examined in the current paper
and listed in Table 1.  Relative sizes and intensities of the images
are not to scale.}
\label{allgals}
\end{figure}

\begin{figure}
\figurenum{2}
\centerline{\psfig{figure=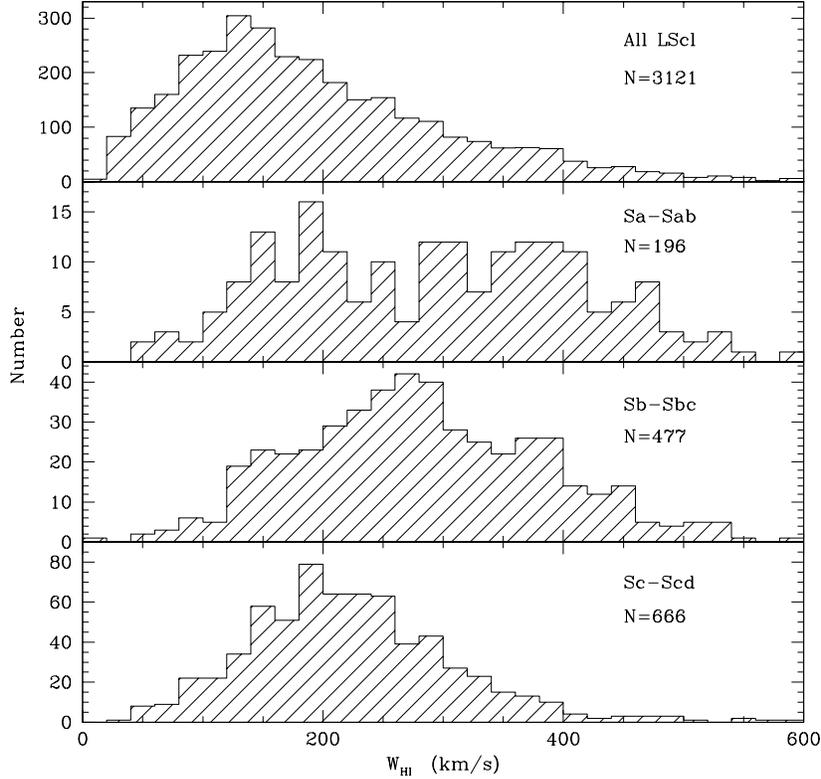,height=4.5in}}
\caption{HI line width distributions for disk galaxies in the Local
Supercluster, by morphological type. Note the small population density with
HI line withs less than about 100~km~s$-1$. 
\label{HIwidths}} 
\end{figure}

\begin{figure}
\figurenum{3}
\caption{Illustration of the method used for quantifying
asymmetry. Shown are galaxies NGC~1637~(left) and
\hbox{MCG-1-60-011~(right)}. The galaxy image is sectioned into some
number of trapezoidal wedges centered on the nucleus in which
photometry is performed. Some galaxies are not precicely face--on; in
these cases, so long as the inclination is $i \lesssim 30\arcdeg$, the
sector pattern may be projected to the inclination of the galaxy, as
is the case for NGC~1637 above.
\label{pizzapie}}
\end{figure}

\begin{figure}
\figurenum{4}
\caption{NGC~5474 with surrounding sector pattern. This galaxy,
because of its extreme asymmetry, was used to test the sensitivity of
the method of sectors. NGC~5474 lies within the gravitational field of
M101, which lies about 1$\arcdeg$ to the NNW.
\label{n5474}}
\end{figure}

\begin{figure}
\figurenum{5}
\centerline{\psfig{figure=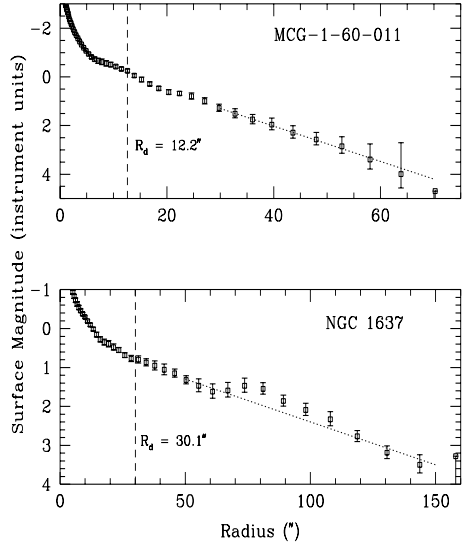,height=3.in,width=2.5in}}
\caption{The relative surface brightness profile, in instrumental
units, for the galaxies \hbox{MCG-1-60-011} (top) and NGC~1637 (bottom). The
dotted line in each panel shows the best linear fit to the outer
portions of the profile. The derived disk scale length, $R_d$, is
indicated by the dashed verticals. Each point is associated with an
elliptical isophote. The average ellipticity of these isophotes over
the region measured with the sector pattern is used to estimate the
inclination of the galaxy.
\label{disks}}
\end{figure}

\begin{figure}
\figurenum{6}
\caption{The galaxy NGC~2326 exhibits a single spiral arm in the outer
regions which can be distinguished for three turns about the
nucleus. Due to its strong $m=1$ spiral mode, such a galaxy would be
detected as lopsided with the methods of RZ and C97, but because its
integrated luminocity is evenly distributed in $\theta$, it is
classified as symmetric by the present method.
\label{n2326}}
\end{figure}


\begin{references}

\reference{bls}Baldwin, J.E., Lynden-Bell, D., \& Sancisi,
R. 1980, \mnras, 193, 313. (BLS)
\reference{bdv}Binney, J., \& de Vaucouleurs, G. 1981, \mnras, 194,
679.
\reference{block}Block, D.L., Bertin, G., Stockton, A., Grosbol, P., Moorwood,
A.F.M., \& Peletier, R.F. 1994, \aap, 288, 365.
\reference{b92}Broeils, A. 1992, Ph.D. Thesis, Univ. of Groningen.
\reference{c97}Conselice, C.J. 1997, \pasp, 109, 1251. (C97)
\reference{rc3} de Vaucouleurs, G., de Vaucouleurs, A., Corwin, H.G.,
Buta, R.J., \& Paturel, G., 1991. {\it Third Reference Catalogue of Bright
Galaxies\/} (RC3), University of Texas Press, Austin.
\reference{ee95}Elmegreen, D.M., \& Elmegreen, B.G. 1995, \apj, 445, 591.
\reference{rg98}Giovanelli, R., Dale, D.A., Haynes, M.P., \& Hardy,
E. 1998, in preparation.
\reference{h98a}Haynes, M.P., Hogg, D.E., Maddalena, R.J., Roberts,
M.S., \& van Zee, L. 1998a, \aj, 115, 62.
\reference{h98b}Haynes, M.P., Giovanelli, R., Salzer, J.J.,  Wegner,
G.,  Freudling, W.F., da Costa, L.N., Herter, T. \& Vogt, N.P. 1998b,
in preparation.
\reference{h98c}Haynes, M.P., Giovanelli, R., Chamaraux, P. da Costa,
L.N., Freudling, W.F., Salzer, J.J. \& Wegner, G. 1998c, in
preparation.
\reference{jog}Jog, C.J. 1997, \apj, 488, 642.
\reference{jc}Junqueira, S., \& Combes, F. 1996, \aap, 312, 703.
\reference{EU}Lauberts, A. 1982. {\it The ESO/Uppsala Survey of the
ESO (B) Atlas.\/} European Southern Observatory, Munchen.
\reference{ls98}Levine, S.E., \& Sparke, L.S. 1998, \apjl, 496, L13.
\reference{l97}Lewis, B.M. 1987, \apjs, 63, 515
\reference{ugc}Nilson, P. 1973. {\it Uppsala General Catalogue of
Galaxies.\/} Uppsala Astron. Obs., Uppsala.
\reference{rz95}Rix, H.-W., \& Zaritsky, D. 1995, \apj, 447, 82 (RZ).
\reference{rs}Richter, O.-G., \& Sancisi, R. 1994, \aap, 290, L9.
\reference{rh94}Roberts, M.S., \& Haynes, M.P. 1994, \araa, 32, 115.
\reference{rownd}Rownd, B.K., Dickey, J.M., \& Helou, G. 1994, \aj,
108, 1638.
\reference{sfz}Schoenmakers, R.H.M., Franx, M., \& de Zeeuw,
P.T. 1997, \mnras, 292, 349.
\reference{st96}Syer, D., \& Tremaine, S. 1996, \mnras, 281, 925.
\reference{vdHH}van der Hulst, J.M. \& Huchtmeier, W.K. 1979, \aap,
78, 82.
\reference{mcg}Vorontsov-Velyaminov, B.A., Archipova, V.P., \&
Krasnogorskaja, A.A. 1968. {\it Morphological Catalogue of
Galaxies.\/} Moscow State Univ., Moscow.
\reference{rz97}Zaritsky, D., \& Rix, H.-W. 1997, \apj, 477, 118 (ZR).
\reference{cgcg}Zwicky, F., Herzog, E., Kowal, C.T., Wild, P., \&
Karpowicz, M. 1960. {\it Catalogue of Galaxies and Clusters of
Galaxies.\/} Calif. Inst. Tech., Pasadena. (six volumes)

\end{references}
\end{document}